\documentclass[aps,twocolumn,showpacs,floatfix]{revtex4}
\usepackage{amsmath}
\usepackage{amsfonts}
\usepackage{amssymb}
\usepackage{epsfig}
\usepackage{graphicx}

\begin{document}
\title{Reactive-Coupling-Induced Normal Mode Splittings in Microdisk Resonators Coupled to Waveguides}
\author{Sumei Huang and G. S. Agarwal}
\affiliation{Department of Physics, Oklahoma State University,
Stillwater, Oklahoma 74078, USA}

\date{\today}

\begin{abstract}
We study the optomechanical design introduced by M. Li \textit{et al.} [Phys.
Rev. Lett. {\bf 103}, 223901 (2009)], which is very effective for
investigations of the effects of reactive coupling. We show the normal
mode splitting that is due solely to reactive coupling rather than
due to dispersive coupling. We suggest feeding the waveguide with a
pump field along with a probe field and scanning the output probe
for evidence of reactive-coupling-induced normal mode splitting.
\end{abstract}
 \maketitle

\renewcommand{\thesection}{\Roman{section}}
\setcounter{section}{0}
\section{Introduction}
\renewcommand{\baselinestretch}{1}\small\normalsize

In a recent paper Li \textit{et al.} \cite{Tang} presented a new design for an
optomechanical system that consists of a microdisk resonator
coupled to a waveguide. This design has several attractive features. Besides its universality, it enables one to study the reactive
effects \cite{Clerk,Tang} in optomechanical coupling. The origin of the reactive coupling is well explained in Ref. \cite{Li}. Its origin lies in the mechanical motion dependence of the extrinsic losses of the disk resonator. Further phase-dependent gradient forces lead to reactive coupling. Li \textit{et al.} have
also argued that this design is more effective in achieving cooling of the system to its ground state. While cooling is
desirable for studying quantum effects at the macroscopic scale
\cite{Hartmann,Bhattacharya,Bose,Sumei1,Paternostro,Vitali}, we
examine other possibilities, which do not depend on the cooling of
the system, to investigate the effects arising from strong reactive
coupling. Since optomechanical coupling effects are
intrinsically nonlinear, we examine the nonlinear response of the
microdisk resonator to pump probe fields. We report reactive-coupling-induced normal mode splitting. Note that in previous
works \cite{Marquardt,Kippenberg,Aspelmeyer,Sumei2} on normal mode
splitting in optomechanical devices, only dispersive coupling was
used. In this paper, we report on normal mode splitting due to
reactive effects.

The paper is organized as follows. In Sec. II, the physical system is
introduced and the time evolutions of the expectation values of the
system operators are given and solved. In Sec. III, the expectation
value of the output fields is calculated, and the nonlinear
susceptibilities for Stokes and anti-Stokes processes are obtained.
In Sec. IV, we discuss normal mode splitting in output fields
with or without reactive coupling. We find that there is no normal mode
splitting in output fields in the absence of reactive
coupling. However, normal mode splitting occurs in output
fields in the presence of reactive coupling.
\section{Model}
\begin{figure}[htp]
 \scalebox{0.8}{\includegraphics{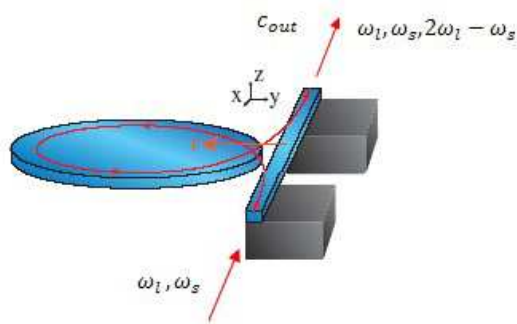}}% Here is how to import EPS art
 \caption{\label{Fig1}(Color online)  Sketch of the studied system (from Ref.\cite{Tang}). The microdisk cavity is driven by a pump field and a Stokes field. The nonlinearity of the interaction also generates anti-Stokes field.}
\end{figure}

 We consider the system shown in Fig.~\ref{Fig1}, in which a microdisk cavity is coupled to a freestanding waveguide. A strong pump field with frequency
$\omega_{l}$ and a weak Stokes field with frequency $\omega_{s}$
enter the system through the waveguide. The waveguide will move
along the $y$ direction under the action of the optical force exerted
by the photons from the cavity. Further, considering the dispersive
coupling and reactive coupling between the waveguide and the cavity,
displacement $q$ of the waveguide from its equilibrium position
will change the resonant frequency of the cavity field and the
cavity decay rate, represented by $\omega_{c}(q)$
 and $\kappa_{e}(q)$, respectively.

In a rotating frame at pump frequency $\omega_{l}$, the
Hamiltonian of the system is given by \cite{Tang}
\begin{equation}\label{1}
\begin{array}{lcl}
\displaystyle H=\hbar
[\omega_{c}(q)-\omega_{l}]c^{\dag}c+\frac{p^2}{2m}+\frac{1}{2}m\omega_{m}^2q^2\vspace{0.1in}\\\hspace{0.3in}+\hbar\frac{
L}{c}\tilde{n}_{g}(\omega_{l}\varepsilon_{l}^2+\omega_{s}|\varepsilon_{s}|^2)+i\hbar\sqrt{2\kappa_{e}(q)}\varepsilon_{l}(c^{\dag}-c)\vspace{0.1in}\\\hspace{0.3in}
+i\hbar\sqrt{2\kappa_{e}(q)}(\varepsilon_{s}e^{-i\delta t}c^{\dag}-\varepsilon^{*}_{s}e^{i\delta t}c).
\end{array}
\end{equation}
The first term is the energy of the cavity field, whose annihilation
(creation) operators are denoted $c(c^{\dag})$. The second and
third terms are the energy of the waveguide with mass $m$, frequency
$\omega_{m}$, and momentum operator $p$. The fourth term gives the
interactions between the waveguide and the incident fields (the pump
field and the Stokes field), $L$ is the length of the waveguide, $c$
is the speed of light in vacuum, $\tilde{n}_{g}$ is the group index
of the waveguide optical mode \cite{Pernice}, $\varepsilon_{l}$ and
$|\varepsilon_{s}|$ are the amplitudes of the pump field and the
Stokes field, respectively, and they are related to their
corresponding power $\wp_{l}$ and $\wp_{s}$ by
$\varepsilon_{l}=\displaystyle\sqrt{\frac{\wp_{l}}{\hbar\omega_{l}}}$
and
$|\varepsilon_{s}|=\displaystyle\sqrt{\frac{\wp_{s}}{\hbar\omega_{s}}}$.
The latter two terms describe the coupling of the cavity field to the
pump field and the Stokes field, respectively. And $\delta=\omega_{s}-\omega_{l}$ is the detuning between the Stokes field and the pump field. We would study the physical effects by scanning the Stokes laser.

For a small displacement $q$, $\omega_{c}(q)$ and $\kappa_{e}(q)$
can be expanded to the first order of $q$,
\begin{equation}\label{2}
\begin{array}{lcl}
\displaystyle\omega_{c}(q)\approx\omega_{c}+q\chi,\vspace{0.2in}\\
\displaystyle\kappa_{e}(q)\approx\kappa_{e}+q\kappa_{om},
\end{array}
\end{equation}
thus the quantities $\chi$ and $\kappa_{om}$ describe the
cavity-waveguide dispersive and reactive coupling strength,
respectively. Further, note that the photons in the cavity can leak
out of the cavity by an intrinsic damping rate $\kappa_{i}$ of the
cavity and by a rate of $\kappa_{e}(q)$ due to the reactive coupling
between the waveguide and the cavity. In addition, the velocity of
the waveguide is damped at a rate of $\gamma_{m}$. Applying the
Heisenberg equation of motion and adding the damping terms, the time
evolutions of the expectation values ($\langle q\rangle,\langle
p\rangle$, and $\langle c\rangle$) for the system can be expressed
as

\begin{equation}\label{3}
\begin{array}{lcl}
\displaystyle \langle\dot{q}\rangle=\frac{\langle
p\rangle}{m},\vspace{0.2in}\\
\displaystyle\langle\dot{p}\rangle=-m\omega_{m}^2\langle
q\rangle-\hbar\chi\langle c^{\dag}\rangle\langle
c\rangle\displaystyle-2\hbar\frac{\kappa_{om}}{\sqrt{\kappa}}\vspace{0.2in}\\\hspace{0.4in}\times\mbox{Im}[(\varepsilon_{l}+\varepsilon^{*}_{s}e^{i\delta t})\langle
c\rangle]-\gamma_{m}\langle p\rangle,\vspace{0.2in}\\
\displaystyle\langle\dot{c}\rangle=-[\kappa+\langle
q\rangle\kappa_{om}+i(\omega_{c}-\omega_{l}+\langle
q\rangle\chi)]\langle
c\rangle\vspace{0.2in}\\\hspace{0.4in}\displaystyle+\sqrt{\kappa}[1+\langle
q\rangle\frac{\kappa_{om}}{\kappa}](\varepsilon_{l}+\varepsilon_{s}e^{-i\delta t}),
\end{array}
\end{equation}
where we have used the mean field assumption $\langle
qc\rangle=\langle q\rangle\langle c\rangle$, expanded
$\kappa_{e}(q)$ to the first order of $q$, and assumed
$\kappa_{e}\approx\kappa_{i}\approx\kappa/2$, where $\kappa$ is the
half-linewidth of the cavity field. It should be noted that the
steady-state solution of Eq. (\ref{3}) contains an infinite number
of frequencies. Since the Stokes field $\varepsilon_{s}$ is much
weaker than the pump field $\varepsilon_{l}$, the steady-state
solution of Eq. (\ref{3}) can be simplified to first order in
$\varepsilon_{s}$ only. We find that in the limit
$t\rightarrow\infty$, each $\langle q\rangle$,$\langle p\rangle$,
and $\langle c\rangle$ has the form
\begin{equation}\label{4}
\begin{array}{lcl}
\langle
s\rangle=s_{0}+s_{+}\varepsilon_{s}e^{-i\delta t}+s_{-}\varepsilon^{*}_{s}e^{i\delta t},
\end{array}
\end{equation}
where $s$ stands for any of the three quantities $q$, $p$, and $c$.
Thus the expectation values $(\langle q\rangle,\langle p\rangle$,
and $\langle c\rangle$) oscillate at three frequencies
($\omega_{l}$, $\omega_{s}$, and 2$\omega_{l}-\omega_{s}$).
Substituting Eq. (\ref{4}) into Eq. (\ref{3}), ignoring those terms
containing the small quantities $\varepsilon_{s}^{2}$,
$\varepsilon_{s}^{*2}$, $|\varepsilon_{s}|^{2}$, and equating
coefficients of terms with the same frequency, respectively, we
obtain the following results
\begin{equation}\label{5}
\begin{array}{lcl}
\displaystyle
c_{0}=\frac{A\varepsilon_{l}}{\kappa+q_{0}\kappa_{om}+i\Delta},\vspace{0.1in}\\
q_{0}=\displaystyle
-\frac{\hbar}{m\omega_{m}^2}[\chi|c_{0}|^2+i\frac{\kappa_{om}}{\sqrt{\kappa}}\varepsilon_{l}(c_{0}^{*}-c_{0})],\vspace{.1in}\\
c_{+}=\displaystyle
\frac{1}{d(\delta)}[A(BE+FJ)-i\hbar\frac{\kappa_{om}}{\sqrt{\kappa}}c_{0}^{*}BF^{*}],\vspace{.1in}\\
c_{-}=\displaystyle
\frac{F^{*}}{d^{*}(\delta)}(-AJ+i\hbar\frac{\kappa_{om}}{\sqrt{\kappa}}c_{0}V),\vspace{0.1in}\\
q_{+}=\displaystyle
\frac{B}{d(\delta)}(-AJ^{*}-i\hbar\frac{\kappa_{om}}{\sqrt{\kappa}}c_{0}^{*}V^{*}),\vspace{0.1in}\\
q_{-}=(q_{+})^{*},
\end{array}
\end{equation}
where
\begin{equation}\label{6}
\Delta=\omega_{c}-\omega_{l}+\chi q_{0},
\end{equation}
\begin{equation}\label{7}
\begin{array}{lcl}
d(\delta)=V^{*}(BE+FJ)+BF^{*}J^{*},
\end{array}
\end{equation}
and
$A=\displaystyle\sqrt{\kappa}(1+\frac{\kappa_{om}}{\kappa}q_{0})$,
$B=\kappa+q_{0}\kappa_{om}-i(\Delta+\delta)$,
$E=m(\omega_{m}^{2}-\delta^{2}-i\gamma_{m}\delta)$,
$F=-c_{0}^{*}(\kappa_{om}-i\chi)+\displaystyle\frac{\kappa_{om}}{\sqrt{\kappa}}\varepsilon_{l}$,
$J=\chi\hbar
c_{0}+i\hbar\displaystyle\frac{\kappa_{om}}{\sqrt{\kappa}}\varepsilon_{l}$,
$V=\kappa+q_{0}\kappa_{om}-i(\Delta-\delta)$. The
approach used in this paper is similar to our earlier work
\cite{Sumei2} which dealt with optomechanical systems with
dispersive coupling only.

\section{output fields}
To investigate the normal mode splitting of the output fields, we need to calculate their expectation value. It can be obtained by using the input-output relation \cite{Walls} $\langle
c_{out}\rangle=\sqrt{2\kappa_{e}(q)}\langle c\rangle$. If we write
$\langle c_{out}\rangle$ as
\begin{equation}\label{8}
\begin{array}{lcl}
\langle
c_{out}\rangle=c_{l}+\varepsilon_{s}e^{-i\delta t}c_{s}+\varepsilon^{*}_{s}e^{i\delta t}c_{as},
\end{array}
\end{equation}
where $c_{l}$ is the response at the pump frequency $\omega_{l}$,
$c_{s}$ is the response at the Stokes frequency $\omega_{s}$, and
$c_{as}$ is the field generated at the new anti-Stokes frequency
$2\omega_{l}-\omega_{s}$. Then we have
\begin{equation}\label{9}
\begin{array}{lcl}
c_{l}=\displaystyle\sqrt{\kappa}(1+\frac{\kappa_{om}}{\kappa}q_{0})c_{0},\vspace{0.1in}\\
c_{s}=\displaystyle\frac{\kappa_{om}}{\sqrt{\kappa}}q_{+}c_{0}+\sqrt{\kappa}(1+\frac{\kappa_{om}}{\kappa}q_{0})c_{+},\vspace{0.1in}\\
c_{as}=\displaystyle\frac{\kappa_{om}}{\sqrt{\kappa}}q_{-}c_{0}+\sqrt{\kappa}(1+\frac{\kappa_{om}}{\kappa}q_{0})c_{-}.\vspace{0.1in}\\
\end{array}
\end{equation}
Furthermore, whether there is normal mode splitting in the output
fields is determined by the roots of the denominator
$d(\delta)$ of $c_{s}$. Here we examine the roots of
$d(\delta)$ given by Eq. (\ref{7}) numerically.
\begin{figure}[htp]
 \scalebox{0.65}{\includegraphics{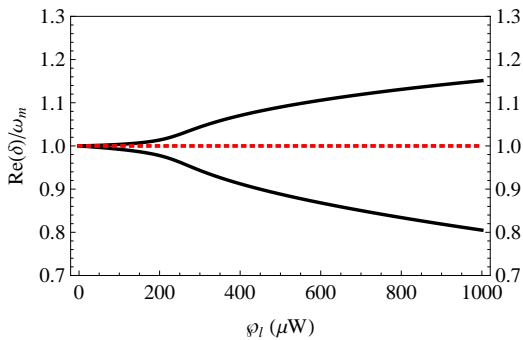}}% Here is how to import EPS art
 \caption{\label{Fig2}(Color online)  The real roots of $d(\delta)$ in the domain
Re$(\delta)>0$ as a function of the pump power
$\wp_{l}$ for $\kappa_{om}=0$ (dotted curve) and
$\kappa_{om}=-2\pi\times26.6$ MHz/nm (solid curve).}
\end{figure}
\begin{figure}[htp]
 \scalebox{0.65}{\includegraphics{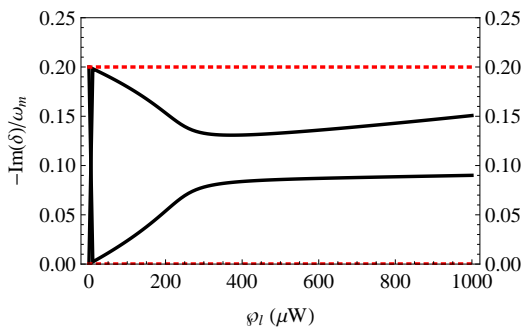}}% Here is how to import EPS art
 \caption{\label{Fig3}(Color online)  Imaginary parts of the roots of $d(\delta)$ as a function of the pump power
$\wp_{l}$ for $\kappa_{om}=0$ (dotted curve) and
$\kappa_{om}=-2\pi\times26.6$ MHz/nm (solid curve).}
\end{figure}

The response of the system is expected to be especially significant
if we choose $\omega_{s}$ corresponding to a sideband
$\omega_{s}=\omega_{l}\pm\omega_{m}$ or
$\omega_{s}=\omega_{l}\pm\Delta$, so we consider the case
$\Delta=\omega_{m}$. The other parameters are chosen from a recent
experiment focusing on the effect of the reactive force on the
waveguide \cite{Tang}: the wavelength of the laser $\lambda=2\pi
c/\omega_l=1564.25$ nm, $\chi=2\pi\times 2$ MHz/nm, $m=2$ pg (density of the silicon waveguide, 2.33 g/cm$^3$; length, 10 $\mu$m;
width, 300 nm; height, 300 nm), $\kappa=0.2 \omega_{m}$,
$\omega_m=2\pi\times25.45$ MHz, and the mechanical quality factor
$Q=\omega_{m}/\gamma_{m}=5000$. In the following, we work in the
 stable regime of the system.

Figure~\ref{Fig2} shows the variation of the real parts of the
roots of $d(\delta)$ in the domain
Re$(\delta)>0$ with increasing pump power for no
reactive coupling, $\kappa_{om}=0$, and for
$\kappa_{om}=-2\pi\times26.6$ MHz/nm. For $\kappa_{om}=0$, the
interaction of the waveguide with the cavity is purely dispersive;
the cavity decay rate does not depend on the displacement of the
waveguide. In this case, the real parts of the roots of
$d(\delta)$ always have two equal values with
increasing pump power. Thus there is no splitting because the
dispersive coupling is not strong enough. However, for
$\kappa_{om}=-2\pi\times26.6$ MHz/nm, the system has both dispersive
and reactive couplings, the cavity decay rate depends on the
displacement of the waveguide, and the real parts of the roots of
$d(\delta)$ will change from two equal values to two
different values with increasing pump power. And the difference
between two real parts of the roots of $d(\delta)$ in
the domain Re$(\delta)>0$ is increased with
increasing pump power. Therefore, the reactive coupling between
the waveguide and the cavity can result in normal mode splitting
of the output fields, and the peak separation becomes larger with
increasing pump power. Figure~\ref{Fig3} shows the variation
of the imaginary parts of the roots of $d(\delta)$
with increasing pump power for zero reactive coupling
$\kappa_{om}=0$ and nonzero reactive coupling
$\kappa_{om}=-2\pi\times26.6$ MHz/nm. For $\kappa_{om}=0$, the
imaginary parts of the roots of $d(\delta)$ do not
change with increasing pump power. However, for
$\kappa_{om}=-2\pi\times26.6$ MHz, the imaginary parts of the roots
of $d(\delta)$ change with increasing pump power.
We thus conclude that for the present microdisk resonator coupled
to a waveguide the normal mode splitting is solely due to the
reactive coupling.

\begin{figure}[!h]
\begin{center}
\scalebox{0.7}{\includegraphics{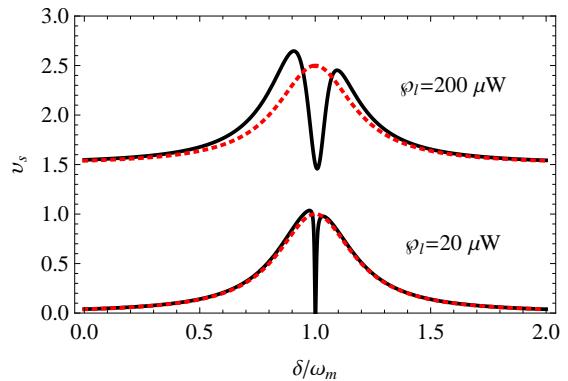}} \caption{\label{Fig4} (Color
online) The lower two curves show the normalized quadrature $v_{s}$ as a
function of the normalized detuning between the Stokes field and the pump field,
$\delta/\omega_{m}$ for $\kappa_{om}=0$ (dotted
curve) and $\kappa_{om}=-2\pi\times26.6$ MHz/nm (solid curve) for
pump power $\wp_{l}=20$ $\mu$W. The upper two curves give the normalized
quadrature $v_{s}$+1.5 for
pump power $\wp_{l}=200$ $\mu$W.}
\end{center}
\end{figure}

\begin{figure}[!h]
\begin{center}
\scalebox{0.7}{\includegraphics{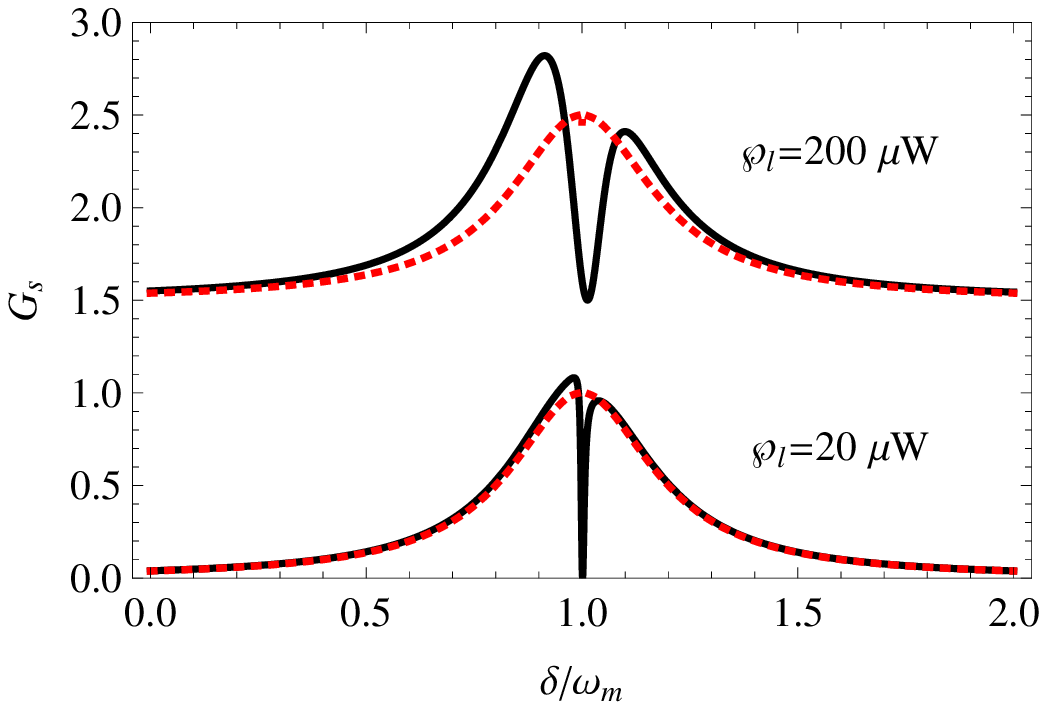}} \caption{\label{Fig5} (Color
online) The lower two curves show the normalized output power $G_{s}$ as
a function of the normalized detuning between the Stokes field and the pump field,
$\delta/\omega_{m}$ for $\kappa_{om}=0$ (dotted
curve) and $\kappa_{om}=-2\pi\times26.6$ MHz/nm (solid curve) for
pump power $\wp_{l}=20$ $\mu$W. The upper two curves give the normalized
output power $G_{s}$+1.5 for
pump power $\wp_{l}=200$ $\mu$W.}
\end{center}
\end{figure}
\begin{figure}[!h]
\begin{center}

\end{center}
\end{figure}
\begin{figure}[!h]
\begin{center}
\scalebox{0.7}{\includegraphics{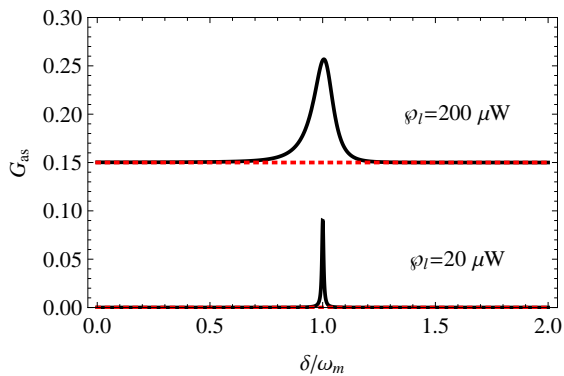}} \caption{\label{Fig6} (Color
online) The lower two curves show the normalized output power $G_{as}$
as a function of the normalized detuning between the Stokes field and the pump field,
$\delta/\omega_{m}$ for $\kappa_{om}=0$ (dotted
curve) and $\kappa_{om}=-2\pi\times26.6$ MHz/nm (solid curve) for
pump power $\wp_{l}=20$ $\mu$W. The upper two curves give the normalized
output power $G_{as}$+0.15 for
pump power $\wp_{l}=200$ $\mu$W. }
\end{center}
\end{figure}
\section{Normal Mode Splitting In Output Fields}
We now discuss how the output fields depend on the behavior of the
roots of $d(\delta)$. For convenience, we normalize all quantities to the input Stokes power
$\wp_{s}$. Assuming that $\varepsilon_{s}$ is real, we express the
output power at the Stokes frequency $\omega_{s}$ in terms of the
input Stokes power
\begin{equation}\label{10}
\begin{array}{lcl}
G_{s}=\displaystyle\frac{\hbar\omega_{s}|\varepsilon_{s}c_{s}|^2}{\wp_s}=|c_{s}|^2.
\end{array}
\end{equation}
Further, we introduce the two quadratures of the Stokes component of
the output fields by
$\displaystyle\upsilon_{s}=\frac{c_{s}+c_{s}^*}{2}$ and
$\displaystyle\tilde{\upsilon}_{s}=\frac{c_{s}-c_{s}^*}{2i}$. One
can measure either the quadratures of the output by homodyne
techniques or the intensity of the output. For brevity, we only show
$\upsilon_{s}$ and $G_{s}$ as a function of the normalized detuning between the Stokes field and the pump field
$\delta/\omega_{m}$ for this model, without reactive
coupling ($\kappa_{om}$=0) and with it
($\kappa_{om}=-2\pi\times26.6$ MHz/nm), for different pump powers in
Figs.~\ref{Fig4}--\ref{Fig5}. For $\kappa_{om}$=0, it is found that
$\upsilon_{s}$ has a Lorentzian lineshape corresponding to the
absorptive behavior. Note that $\upsilon_{s}$ and $G_{s}$ exhibit no
splitting when $\kappa_{om}$=0. However, for
$\kappa_{om}=-2\pi\times26.6$ MHz/nm, it is clearly seen that normal
mode splitting appears in $\upsilon_{s}$ and $G_{s}$. Therefore reactive coupling can lead to the appearance of normal mode
splitting in the output Stokes field.  And the peak separation increases with increasing pump power \cite{Agarwal}. The dip at the line
center exhibits power broadening. We also find that the Stokes field can
be amplified by the stimulated process. Obviously the maximum gain
$G_{s}$ for the Stokes field depends on the system parameters. For a
pump power $\wp_{l}=200$ $\mu$W, the maximum gain for the Stokes
field is about 1.3.

Note that the nonlinear nature of the reactive coupling generates
anti-Stokes radiation. In a similar way, we define a normalized
output power at the anti-Stokes frequency $2\omega_{l}-\omega_{s}$
as
\begin{equation}\label{11}
\begin{array}{lcl}
G_{as}=\displaystyle\frac{\hbar(2\omega_{l}-\omega_{s})|\varepsilon_{s}c_{as}|^2}{\wp_s}=|c_{as}|^2.
\end{array}
\end{equation}
The plots of $G_{as}$ versus the normalized detuning between the Stokes field and the pump field
$\delta/\omega_{m}$ for this model, without reactive
coupling ($\kappa_{om}$=0) and with it
($\kappa_{om}=-2\pi\times26.6$ MHz/nm), for different pump powers are
presented in Fig.~\ref{Fig6}. We can see that $G_{as}\approx0$ for
$\kappa_{om}$=0. The reason is that the dispersive coupling constant
$\chi$ is too small. However, for $\kappa_{om}=-2\pi\times26.6$
MHz/nm, $G_{as}$ is not equal to zero. This shows that the
optomechanical system can generate an anti-Stokes field with
frequency $(2\omega_{l}-\omega_{s})$ due to the reactive coupling.
For pump power $\wp_{l}=200$ $\mu$W, the maximum gain defined with
reference to the input Stokes power for the anti-Stokes field is about
0.1.

\section{Conclusions}
In conclusion, we have observed normal mode splitting of output fields due to reactive coupling between the waveguide and
the cavity. Meanwhile, the separation of the peaks increases for
larger pump powers. Further, the reactive coupling can also cause
four-wave mixing, which creates an anti-Stokes component generated by
the optomechanical system.

We gratefully acknowledge support from NSF Grant No. PHYS
0653494.

\end{document}